\documentclass[12pt]{article}
\pdfoutput=1

\usepackage{amsmath,amssymb,amsfonts,bbm,epsfig,cancel,cite,setspace,bigstrut,framed,tensor,slashed}
\usepackage{color}
\usepackage{pifont}
\usepackage{hyperref}
\usepackage{scalerel}
\usepackage{booktabs}
\usepackage{tikz}

\newcommand\fat[1]{\ThisStyle{\ooalign{%
  \kern.46pt$\SavedStyle#1$\cr\kern.33pt$\SavedStyle#1$\cr%
  \kern.2pt$\SavedStyle#1$\cr$\SavedStyle#1$}}}

\makeatletter \@addtoreset{equation}{section} \makeatother

\newcommand{\SM}{S^\text{M}}
\newcommand{\SNM}{S^\text{NM}}
\newcommand{\tSNM}{\widetilde{S}^\text{NM}}
\DeclareMathOperator{\gh}{gh}

\newcommand{\nn}{\nonumber}
\newcommand{\comment}[1]{}

\newcommand{\bR}{{\mathbf R}}

\newcommand{\ks}{{\mathfrak s}}
\newcommand{\ko}{{\mathfrak o}}
\newcommand{\ku}{{\mathfrak u}}
\newcommand{\kf}{{\mathfrak f}}

\newcommand{\kp}{{\mathfrak p}}
\newcommand{\ke}{{\mathfrak e}}

\newcommand{\kso}{{\ks\ko}}
\newcommand{\ksp}{{\ks\kp}}
\newcommand{\ksu}{{\ks\ku}}
\newcommand{\cN}{{\cal N}}

\newcommand{\cM}{{\cal M}}

\newcommand{\cR}{{\cal R}}

\newcommand{\cD}{{\cal D}}

\newcommand{\cF}{{\cal F}}

\newcommand{\cV}{{\cal V}}

\newcommand{\ii}{\mathbbm{i}}

\newcommand{\ch}{{\rm ch}}

\newcommand{\bi}{\begin{itemize}}
\newcommand{\ei}{\end{itemize}}
\newcommand{\beq}{\begin{equation}}
\newcommand{\eeq}{\end{equation}}
\newcommand{\bea}{\begin{eqnarray}}
\newcommand{\eea}{\end{eqnarray}}

\makeatletter
\newsavebox{\@brx}
\newcommand{\llangle}[1][]{\savebox{\@brx}{\(\m@th{#1\langle}\)}%
  \mathopen{\copy\@brx\kern-0.5\wd\@brx\usebox{\@brx}}}
\newcommand{\rrangle}[1][]{\savebox{\@brx}{\(\m@th{#1\rangle}\)}%
  \mathclose{\copy\@brx\kern-0.5\wd\@brx\usebox{\@brx}}}
\makeatother

\begin{document}

\begin{titlepage}
\vfill
\begin{flushright}
{\tt\normalsize KIAS-P24005}\\

\end{flushright}
\vfill

\begin{center}
{\Large\bf Axial Anomalies of  \\ Maximally Supersymmetric Tensor Theories}

\vskip 1.5cm

Piljin Yi$\,^1$, Yi Zhang$\,^2$
\vskip 5mm

{\it $^1$ School of Physics,
Korea Institute for Advanced Study, Seoul 02455, Korea\\}
\smallskip
{\it  $^2$ Center for High Energy Physics, Peking University, Beijing 100871, China}

\end{center}
\vfill

\begin{abstract}

We revisit anomalies of $(4,0)$ and $(3,1)$ maximally supersymmetric tensor theories in $d=6$. A $(4,0)$ on-shell tensor multiplet descends to that of the $d=5$ maximal supergravity upon a dimensional reduction, hypothesized to offer a strong-coupled UV completion of the latter in the same sense of $(2,0)$ theories as the UV completion of $d=5$
$\cN=2$ pure Yang-Mills. The gravitational anomalies, found to be nonvanishing, had been computed, although its relevance in the absence of the $d=6$ metric is not obvious. We perform a comprehensive anomaly computation for $(4,0)$ and $(3,1)$ tensor supermultiplets, respectively, for $Sp(4)$ and $Sp(3)\times Sp(1)$ $R$-symmetry anomalies
and the mixed $R$-gravitational anomaly thereof, and find that anomalies
involving $R$-symmetries cancel out identically. We close with questions on how
to address the anomaly in this class of theories with no general covariance. 

\end{abstract}

\vfill
\end{titlepage}

\section{Maximally Supersymmetric Tensor Theories}

In $d=6$, we encounter several interesting tensorial theories, which have
led to many fruitful investigations of the landscape of supersymmetric field
theories in both $d=5$ and $d=6$. In particular, the relation between
$(2,0)$ theories with $d=5$ maximally supersymmetric Yang-Mills theories
is one of the more remarkable findings that string theory gave us in the
past decades. Relative to this, less understood class of tensorial
theories in $d=6$ are those with even larger supersymmetries, $(4,0)$ and
$(3,1)$, of which very little beyond their respective field content \cite{Strathdee:1986jr} is
known.

The $(4,0)$ case is particularly interesting because its on-shell
field content matches that of $d=5$ $\cN=4$ supergravity multiplet
verbatim if we embed the $SO(3)$ little group of the latter to the
former's counterpart $SO(4)\simeq SO(3)\times SO(3)$ diagonally.
The match occurs at the level of the individual Lorentz
multiplets when we compactify a theory on a small circle, which is pretty 
unusual. Since something similar  happens between $d=6$ $(2,0)$
tensor multiplet and $d=5$ $\cN=2$ vector multiplet, this analogy has 
led in the past to a proposal \cite{Hull:2000zn} that perhaps the $(4,0)$ 
theory offers a strong-coupling completion of $d=5$ maximal supergravity
in the same sense that $(2,0)$ theories do for $d=5$ $\cN=2$ Yang-Mills theories.

The field content of $(3,1)$ also descends to that of $d=5$ and $\cN=4$ 
supergravity, after some reorganization given how some $d=6$ Lorentz 
multiplets take nontrivial representations under both factors of $SO(3)\times SO(3)$ 
of the little group. On-shell content of the $(4,0)$ theory are all singlets under one 
$SO(3)$, on the contrary, allowing the simple descent to the $SO(3)$ little group 
of $d=5$ field by field.  Also the $d=5$ $\cN=4$ $R$-symmetry $Sp(4)$ coincides precisely with 
that of the $(4,0)$ theory.

The main technical hurdle against investigating such a claim, or
in fact studying the tensorial theory by itself, is how poorly we
understand the interaction among the chiral fields therein.
Since the metric is not part of these tensor supermultiplets, the
general covariance is replaced by a sequence of tensorial gauge symmetries,
and there remains much uncertainty on how one might go about building
an interacting $d=6$ Lagrangian (See~\cite{Henneaux:2017xsb, Henneaux:2018rub,Bertrand:2020nob,Bertrand:2022pyi} for considerations of free actions). One may suggest that a non-Lagrangian
conformal theory is needed; although this may prove to be the case,
the idea  by itself does not offer practical tools for concrete investigations.

One universal aspect of quantum theories for which the complete detail 
of interactions are not needed is the anomaly. There are three classes 
of perturbative anomalies one might compute: gravitational anomalies, 
$R$-symmetry anomalies, and those of the gauge symmetries of the 
chiral tensors. The anomaly of the last is quite essential 
if one is to investigate the above 5d/6d connection, since the 
$d=5$ general covariance has to descend from these tensorial 
gauge symmetries, given how the $d=5$ metric arises
from a $d=6$ chiral tensor.

The $(4,0)$ and $(3,1)$ superalgebras are both maximal
such in $d=6$. With usual maximally supersymmetric gravitational theories,
all of their standard anomaly polynomials either cancel away or a simple
Green-Schwarz type topological term appears and deals with
the remainder. Examples of the latter can be found in $d=8$ 
\cite{Gaberdiel:1998ui,Minasian:2016hoh,Lao:2023tuj}, while the
cancellation of the anomaly polynomial is a matter of combinatorics 
for $d= 4, 6,10$ \cite{Alvarez-Gaume:1983ihn,Marcus:1985yy}. This
past experience led to a previous investigation of the gravitational
anomalies of $(4,0)$ and $(3,1)$ tensor multiplets \cite{Minasian:2020vxn},
which in the end found  nonvanishing results. However, the latter does not need to discourage us given how 
neither the proposed 5d/6d connection nor the on-shell field content in $d=6$ side 
implies $d=6$ general covariance in the conventional sense.

On the other hand,  one of the cornerstones of the 5d/6d proposal is that
the $d=6$ $(4,0)$ $R$-symmetry, namely $Sp(4)$\footnote{We denote by $Sp(r)$ the compact symplectic group of rank $r$ and $\ksp(r)$ represents the corresponding Lie algebra.} descends unmodified to that
of $d=5$ $\cN=4$. Unlike the $d=5$ diffeomorphisms, these $R$-symmetries
would uplift to $d=6$ verbatim straightforwardly, so there is no room 
to wiggle out if the $R$-symmetry proves to be anomalous. Although we are now familiar with 
anomaly inflow when we embed a field theory to string theory \cite{Witten:1996hc,Freed:1998tg,Green:1996dd,Cheung:1997az,Minasian:1997mm,Kim:2012wc},
this does not seem to happen for maximally supersymmetric models.

As such, the main purpose of this note is to check if indeed $R$-symmetry
anomalies cancel away unlike their gravitational counterpart.
Recall how the anomaly polynomial from a chiral field has the general structure
of index density,
\bea \label{eq:anomalyindexdensity}
\mathbb A(\cR)\wedge \ch_{\bR_G}(\cR)\wedge \ch_{\bR_{R}}(\cF)\ ,
\eea
where $\bR_G$ is a collection of $\kso(d)$ representations determined by
the chiral field in question while $\bR_{R}$ is the relevant representation under the $R$-symmetry.
As usual, $\mathbb A(\cR)$ is the A-roof genus of the tangent bundle while ${\rm ch}_\bR$ means
the Chern character of $\cF$ in the representation $\bR$.

The procedure that computes $\bR_G$'s for various chiral fields in the on-shell multiplet amounts to understanding off-shell field contents, both bosonic and fermionic and the elaborate  gauge fixing thereof. In the next section, we briefly review the result from recent literature \cite{Minasian:2020vxn, Lekeu:2021oti} for this preliminary step.
These contributions would compute the anomaly all the same
if we couple all fields to external gravity and $R$-gauge fields.
In the rest of this note, we will evaluate the anomaly polynomial
and sum over the Lorentz multiplets in the respective tensor multiplets.

In our notations, the on-shell field content of these tensor multiplets are as follows, classified by the little group and the $R$-symmetry group displayed in the respective first lines \cite{Hull:2000zn, Strathdee:1986jr}:
\begin{itemize}
\item $d=6$ $\cN=(4,0)$ $\kso(4)\oplus \ksp(4)$
$$(\mathbf 1, \mathbf 1 ; \mathbf{42} )+ (\mathbf 2, \mathbf 1 ; \mathbf{48} )+(\mathbf 3, \mathbf 1; \mathbf{27})+ (\mathbf 4, \mathbf 1; \mathbf 8)+
(\mathbf 5, \mathbf 1; \mathbf 1)$$
\item $d=6$ $\cN=(3,1)$ $\kso(4)\oplus \ksp(3)\oplus\ksp(1)$
$$(\mathbf 1, \mathbf 1; \mathbf{14}', \mathbf 2)+ (\mathbf 2, \mathbf 1; \mathbf{14}, \mathbf 2)+ (\mathbf 1, \mathbf 2; \mathbf{14}', \mathbf 1)+(\mathbf 3, \mathbf 1; \mathbf 6, \mathbf 2)$$
$$+\;(\mathbf 2, \mathbf 2; \mathbf{14}, \mathbf 1)+ (\mathbf 3, \mathbf 2; \mathbf 6, \mathbf 1)+
(\mathbf 4, \mathbf 1; \mathbf 1, \mathbf 2)+ (\mathbf 4, \mathbf 2; \mathbf 1, \mathbf 1)$$
\end{itemize}
For the $(3,1)$ multiplet, $\mathbf{14}$ is the rank 2 antisymmetric tensor $\fat[2\fat]$ of $\ksp(3)$
and $\mathbf{14}'$ is the rank 3 antisymmetric tensor $\fat[3\fat]$ of $\ksp(3)$, respectively.
For the former,  $\mathbf{27}$ is $\fat[2\fat]$ of $\ksp(4)$ and $\mathbf{48}$ is $\fat[3\fat]$ of $\ksp(4)$.

\section{Off-Shell Field Content}

Under the little algebra $\kso(4) = \ksu(2)\oplus \ksu(2)$, many of the lower dimensional
representations are already well-understood off-shell.  $({\bf 3,1})$ is for example
from a self-dual tensor $B_{\mu\nu}$. Relatively exotic are the fermionic tensor $({\bf 4,1})$ and the bosonic tensors $({\bf 4,2})$ and $({\bf 5,1})$. As for free fermionic exotic tensor $({\bf 4,1})$, the Batalin-Vilkovisky (BV) formulation \cite{Batalin:1981jr, Batalin:1984jr} providing the covariant path integral quantization has been studied in \cite{Lekeu:2021oti}, and we borrow their results for the sake of completeness. A brief review of BV formalism (for first-stage reducible theories) itself is also offered in appendix~\ref{app:BV}.

\subsubsection*{The (4,1) Tensor}

The chiral field $(\mathbf 4,\mathbf 1)$, which we also refer to as the exotic gravitino, has a Lorentz covariant field realization as a chiral fermionic two form $\psi_{\mu\nu}$, satisfying
\begin{equation}
\begin{aligned}
\psi_{\mu\nu} &= -\psi_{\nu\mu}\ , \\ \nonumber
\Gamma_7 \psi_{\mu\nu} &= \psi_{\mu\nu}\, , \nonumber
\end{aligned}
\end{equation}
here $\mu$, $\nu$ are spacetime indices and $\Gamma_7$ is the chirality matrix in $6d$. The field equation leads to $(\mathbf 4,\mathbf 1)$ is\footnote{The field strength $H_{\mu\nu\rho}:= 3 \partial_{[\mu}\psi_{\nu\rho]}$ is self-dual $H=\star H$ under the Hodge star operator. One can derive the self-dual condition using the field equation \eqref{eq:eom} together with the fact that the tensor spinor is chiral. In general, self-duality is weaker than field equation for chiral tensor spinors. See \cite{Henneaux:2017xsb} and \cite{Zhang:2021swj} for proofs.}
\begin{equation} \label{eq:eom}
\Gamma^{\alpha\beta\mu\nu\rho}\partial_{\mu}\psi_{\nu\rho}=0 \,.
\end{equation}
It is straightforward to obtain this equation from a generalized Rarita-Schwinger action 
\begin{equation} \label{Rarita-Schwinger}
S[\psi_{\mu\nu}] = \int d^6x \; \bar{\psi}_{\mu\nu} \Gamma^{\mu\nu\rho\sigma\tau}\partial_{\rho}\psi_{\sigma\tau} \,.
\end{equation}
Both the field equation and the action are invariant under the gauge transformations
\begin{equation} \label{eq:gaugetranformationpsimu}
\delta \psi_{\mu\nu} = 2\, \partial_{[\mu} \epsilon_{\nu]}\,,
\end{equation}
with spacetime dependent vector-spinor parameter $\epsilon_{\nu}$.
This type of gauge symmetry is called \emph{reducible} \cite{Batalin:1984jr}, since there are gauge for gauge transformations
\begin{equation}
\delta \epsilon_\mu = \partial_\mu \lambda\ ,
\end{equation}
indicating that the gauge transformations \eqref{eq:gaugetranformationpsimu} are not linearly independent.
The gauge parameter $\epsilon$ has $n = 6 \times s$ components $\lambda$ has $m= s$ components, for a total of $n-m=(6-1) \times s$ independent gauge symmetries, where $s$ is the dimension of the spinor representation at hand. We had omitted the spinor index $\alpha$ for the fields above. 

This reducibility of gauge transformations complicates the covariant path integral quantization. Thanks to the Batalin–Vilkovisky (BV) field-antifield formalism, one can perform the covariant quantization systematically.

A crucial step in the BV computations is to use a \emph{gauge-fixing fermion}\footnote{This type \eqref{eq:psidelta} of $\Psi$ is called the delta-function gauge-fixing fermion. It is not the only possibility and another admissible type is called the Gaussian gauge-fixing fermion. The dynamical ghosts spectrum resulting from the two  are different but they lead to the same degrees of freedom counting and as well as same anomaly result \cite{Lekeu:2021oti}.}
\begin{align}\label{eq:psidelta}
    \Psi &= \int \! d^6\!x\, \left( \bar{C}'_\mu\, \chi^\mu(\psi) + \ldots \right)\, ,
\end{align}
with the redundant gauge condition
\begin{align}\label{eq:gaugecond2form}
    \chi_\mu(\psi) &\equiv \gamma^\nu \psi_{\mu\nu} - \frac{1}{6-2} \gamma_{\mu\nu\rho} \psi^{\nu\rho} \nn \\
    &= 0 \, ,
\end{align}
which satisfies the constraint
\begin{equation}
    \gamma^\mu \chi_\mu(\psi) = 0\ ,
\end{equation}
identically and hence gives the correct number $n-m=(6-1) \times s$ of gauge conditions to fix the independent gauge transformations. 

We sketch the BV-quantisation process for $\psi_{\mu\nu}$ here:
\begin{itemize}
    \item We introduce ghost fields $\{C^\alpha_\mu, \, c^\alpha \}$ for gauge parameters $\epsilon^\alpha_\mu$ and $\lambda^\alpha$, and then introduce for all the fields $\Phi^I$ at hand their antifields $\Phi^*_I$ to get the minimal sector
    \begin{equation}
    \{ \psi^\alpha_{\mu\nu}\, , C^\alpha_\mu\, , c^\alpha\, , \psi_\alpha^{*\mu\nu}\, , C^{*\mu}_\alpha\, , c^*_\alpha \}\, ,
\end{equation}
together with an extended action called the \emph{minimal action} $S^{\text{M}}$ = $S^{\text{M}}[\Phi, \Phi^*]$. The minimal action is completely determined by a \emph{master equation} $(S^{\text{M}},S^{\text{M}})=0$ and the boundary conditions
\begin{equation}
S[\psi_{\mu\nu}] = S^{\text{M}}[\Phi, \Phi^* = 0]\, .
\end{equation}
    \item For gauge-fixing, one further introduces trivial pairs of auxiliary along with their antifields. The minimal action $S^{\text{M}}$ is then extended to a non-minimal action $S^{\text{NM}}$ which still satisfies the master equation. The antifields are then eliminated according to the formula
\begin{equation}\label{eq:gffermion}
\Phi^*_I = \frac{\delta \Psi}{\delta \Phi^I}\, ,
\end{equation}
where $\Psi(\Phi)$ is an odd functional of ghost number $-1$ depending on the fields only, and hence called the gauge-fixing fermion. If $\Psi$ is well-chosen as in \eqref{eq:psidelta}, the resulting action is properly gauge-fixed and possesses well-defined propagators.
\item A gauge-fixed action read \begin{align}
    S^{\text{gauge-fixed}} = \int \! d^6\!x\, \Big( &-\frac{1}{2}\bar{\hat{\psi}}^{\mu\nu} \slashed{\partial} \hat{\psi}_{\mu\nu} - \frac{1}{2}\bar{\rho} \slashed{\partial} \rho + \bar{C}'^{\mu} \slashed{\partial} C_\mu + \bar{c}' \slashed{\partial} c \nonumber \\
    &+ \bar{d}^\mu \gamma^\nu \hat{\psi}_{\mu\nu} + \bar{\pi} \gamma^\mu C_\mu + \bar{\pi}' \gamma^\mu C'_\mu + \text{c.c.} \Big) \, ,\label{eq:Sgf-twoform-delta-final}
\end{align}
with different trace components of $\psi_{\mu\nu}$ appear:
\begin{equation}\label{eq:decomppsi}
    \psi_{\mu\nu} = \hat{\psi}_{\mu\nu} + ( \gamma_\mu \sigma_\nu - \gamma_\nu \sigma_\mu ) + \gamma_{\mu\nu} \rho
\end{equation}
where $\hat{\psi}_{\mu\nu}$ and $\sigma_\mu$ are gamma-traceless, $\gamma^\nu \hat{\psi}_{\mu\nu} = 0 = \gamma^\mu \sigma_\mu$. The redundant gauge condition \eqref{eq:gaugecond2form} $\chi_\mu(\psi)=0$ is equivalent to $\sigma_\mu=0$ and we then use an auxiliary field $d_\mu$ to impose the gamma-tracelessness of $\hat{\psi}_{\mu\nu}$. Chirality and statistics of all the fields are displayed in table \ref{tab:2formchiralities}.
\begin{table}
\centering
\begin{tabular}{c|cccccccccccc}
& $\psi_{\mu\nu}$ & $\hat{\psi}_{\mu\nu}$ & $\rho$ & $C_\mu$ & $C'_\mu$ & $c$ & $c'$ &  $d_\mu$ & $\pi$ & $\pi'$  \\[\defaultaddspace] \midrule
Chirality & $+$ & $+$ & $+$ & $+$ & $+$ & $+$ & $+$ & $+$ & $+$ & $+$  \\
Grassmann parity & 1 & 1 & 1 & 0 & 0 & 1 & 1 &  1 & 0 & 0
\end{tabular}
\caption{Chirality and Grassmann parity of the various fields appearing in the gauge-fixed action \eqref{eq:Sgf-twoform-delta-final}.}
\label{tab:2formchiralities}
\end{table}
\end{itemize}
Based on the above data, we can now proceed to read off the representations $\bR_G$ of the fields in \eqref{eq:Sgf-twoform-delta-final}.
The relevant path integral measure is (omitting the complex conjugates and auxiliary fields are integrated out)
\begin{equation}\label{eq:measuredelta}
    \int \cD \hat{\psi}_{\mu\nu} \,\cD \rho \,\cD \hat{C}_\mu \, 
    \cD \hat{C}'_\mu \,\cD c \,\cD c'\ ,
\end{equation}
where a `hat' denotes a gamma-traceless field. We use $\mathcal{C}^{\infty}(V)$ to denote set of smooth sections of vector bundles $V$.
\begin{itemize}
    \item The field $\hat{\psi}_{\mu\nu}$ for example can be seen as an element of the formal difference
\begin{equation}\label{eq:psihatcomplex}
    \mathcal{C}^{\infty}(S^+ \otimes \Lambda^2 T^*\!\cM - S^- \otimes T^*\!\cM)\, ,
\end{equation}
 and it contributes as \begin{equation}
\mathcal{C}^{\infty}(S^+ \otimes [\Lambda^2 T^*\!\cM + T^*\!\cM])\, .
\end{equation}
 \item $\hat{C}_\mu$ and $\hat{C}'_\mu$ are with the wrong spin-statistics, so they contribute with a minus sign. For each of them, we have
\begin{equation}
    - \mathcal{C}^{\infty}(S^+ \otimes T^*\!\cM - S^-) = \mathcal{C}^{\infty}(S^+ \otimes [ - T^*\!\cM - 1])\, .
  \end{equation}
    \item  The other dynamical fields $c$, $c'$ and $\rho$ are just ordinary chiral fermions they together contribute to
\begin{equation}
     3 \, \mathcal{C}^{\infty}(S^+) \, .
\end{equation}
\end{itemize}
The effective complex $\mathcal{C}^{\infty} (S^+ \otimes \cV)$ on which the Dirac operator acts in this case is then
\begin{align}
\mathcal{C}^{\infty} (S^+ \otimes \cV) &= \mathcal{C}^{\infty}(S^+ \otimes [\Lambda^2 T^*\!\cM + T^*\!\cM]) + 2 \mathcal{C}^{\infty}(S^+ \otimes [ - T^*\!\cM - 1])  + 3  \mathcal{C}^{\infty}(S^+) \nn\\
    &= \mathcal{C}^{\infty}(S^+ \otimes [\Lambda^2 T^*\!\cM - T^*\!\cM + 1])\ ,\nn
\end{align}
and we identify
\begin{equation}
    \cV = \Lambda^2 T^*\!\cM - T^*\!\cM + 1\, .
\end{equation}
In other words, the above formal collection of representations contributes the factor
\begin{equation}
    \ch_{\bR_G}(\cR)=\ch_{\fat[2\fat]}(\cR)-  \ch_{\rm def}(\cR)+1
\end{equation}
in the index density~\eqref{eq:anomalyindexdensity}.

\subsubsection*{The (4,2) Tensor}

Now we move to discuss the index density contribution from the bosonic fields. 
In the classical paper \cite{Alvarez-Gaume:1983ihn}, the gravitational anomaly for a $2k$-form potential $A$ with self-dual $2k+1$-form field strength $F$ in $d=4k+2$ dimensions is computed.
The trick is to consider generic tensor fields $A$ and $F$ (without duality constraint) as independent variables, and one integrates over both of them in the path-integral using a first-order action.  
In the path-integral measure, the self-dual $F^+$ part and anti-self-dual part $F^-$ both appear. But one can extract anomaly for the (anti-)self-dual part alone as the Jacobian generated by it under transformations of the Lorentz group in the corresponding (anti-)self-dual representation. Since there is no gauge freedom in $F^+$ or $F^-$, there is no need to subtract ghost contributions. Similar computations are already performed in a recent article \cite{Minasian:2020vxn}. We follow the notations of \cite{Minasian:2020vxn} and review the results.

The bosonic exotic field $(\mathbf 4,\mathbf 2)$ has a three-index covariant gauge potential realization \cite{Hull:2000zn} $D_{\mu\nu\rho}$ which satisfies
\begin{equation}
\label{eq:exotic-D}
D_{\mu\nu\rho}=D_{[\mu\nu]\rho}, \quad D_{[\mu\nu\rho]}=0 \, .
\end{equation}
Its field strength is defined\footnote{The comma in the subscript denotes a partial derivative.} as $S_{\mu\nu\rho\sigma\kappa}=\partial_{[\mu} D_{\nu \rho][\sigma, \kappa]}$ which is subject to
one-side self-dual\footnote{Here we are writing in the signature $(-,+,+,+,+,+)$, since the self-duality and the reality must be
imposed in the Lorentzian signature, although the anomaly polynomials are often computed using the Euclidean signature by an analytic contonuation.} 
constraint first three indices 
\begin{equation}
S_{\mu\nu\rho\sigma\kappa}= (\star S)_{\mu\nu\rho\sigma\kappa} =\frac{1}{3!} \epsilon_{\mu\nu\rho \alpha \beta \gamma} {S^{\alpha \beta \gamma}}_{\sigma\kappa} \, . 
\end{equation}
This self-dual constraint is also the field equation for $D$ which leads to the physical degrees of freedom $(\mathbf 4,\mathbf 2)$ in the little group. 

Since the computations are implemented at the level of field strengths, it is convenient to use Dynkin labels of to describe the irreducible representations in which the field strength transforms. The default should be the $\mathfrak{so}(6)$ Dynkin label. But we would like to use the ``A-type'' conventions of $\mathfrak{su}(4)$ (cf. $\mathfrak{su}(4) \cong \mathfrak{so}(6)$),
for example, we call $[0, 1, 0]$ the vector representation and the spinor representation with positive chirality is denoted as $[0, 0, 1]$, while the spinor with negative chirality is represented by $[1, 0, 0]$. 

The self-dual field strength $S$ is in the $[1,0,3]$ representation. 
From the tensor product decomposition 
\begin{equation}
[1,0,1] \otimes [0,0,2] = [1,0,3] \oplus [1,1,1] \oplus [0,0,2] \oplus [0,1,0] \ ,
\end{equation}
and 
\begin{equation}
[1,0,1] \otimes [0,1,0] = [1,1,1] \oplus [0,0,2] \oplus [0,1,0] \oplus [2,0,0] \ ,
\end{equation}
we get 
\begin{equation}
[1,0,3] = \left([1,0,1] \otimes [0,0,2] \right) \ominus \left( [1,0,1] \otimes [0,1,0] \ominus [2,0,0] \right) \,,
\end{equation}
where the symbol $\ominus$ represents a formal subtraction. 
At the level of smooth sections of bundles, we list only the sections with contributions to index densities\footnote{We use $\phi$ to denote 0-forms and $F_3^\pm$ stands for tensor bundle with self-dual (anti-self-dual) three-form fibers.}
\begin{equation}
\begin{aligned}
S \in &\;\mathcal{C}^{\infty}(\Lambda^2 T^*\!\cM \otimes F_3^+) - \mathcal{C}^{\infty}( \Lambda^2 T^*\!\cM \otimes T^*\!\cM) +\mathcal{C}^{\infty}(F_3^-)\\
= &\; \mathcal{C}^{\infty}\left(S^+ \otimes [S^+ \otimes S^+ \otimes S^- - (S^- \otimes T^*\!\cM )^{\oplus 2} - (S^+)^{\oplus 2}]\right).  \\
\end{aligned}
\end{equation}
It follows that the Chern character $\bR_G$ for the $(\mathbf 4,\mathbf 1)$ tensor is 
\begin{equation}
   \ch_{\bR_G}(\cR) = ( \ch_{+ \rm s}(\cR) )^2 \ch_{- \rm s}(\cR)-  2 \ch_{- \rm s}(\cR)  \ch_{\rm def}(\cR) - 2\ch_{+ \rm s}(\cR) \,,
\end{equation}
where we omitted the wedge product symbol between differential forms and $\pm \rm s$ (which we wrote alternatively as $\pm \rm s$ in appendix~\ref{app:A}) stand for the chiral/anti-chiral  spinor representations.
\subsubsection*{The (5,1) Tensor}
 The discussion follows closely the previous paragraph. The covariant field of the $(\mathbf 5,\mathbf 1)$ tensor is represented by $C_{\mu\nu\rho\sigma}$ with the same index symmetries as the Riemann tensor \cite{Hull:2000zn}
\begin{equation} \label{exoticgravitonsymmetry}
C_{\mu\nu\rho\sigma} = C_{\rho\sigma \mu\nu} = C_{[\mu\nu]\rho\sigma} = C_{\mu\nu[\rho\sigma]} \, ,
\end{equation}
\begin{equation}
C_{[\mu\nu\rho]\sigma} = 0 \, .
\end{equation}
The field strength is defined as
\begin{equation} \label{exoticgravitonfieldstrength}
G_{\mu\nu\rho\sigma\tau\kappa} = \partial_{[\mu}C_{\nu\rho][\sigma\tau,\kappa]} \,,
\end{equation}
and the field equation is the double self-dual condition 
 \begin{equation}\label{eq:SD}
\begin{split}
    G_{\mu\nu\rho\sigma\tau\kappa} &= (\star G)_{\mu\nu\rho\sigma\tau\kappa} \equiv \frac{1}{3!} \epsilon_{\mu\nu\rho\alpha\beta\gamma} {G^{\alpha\beta\gamma}}_{\sigma\tau\kappa} \\
   & = (G\star)_{\mu\nu\rho\sigma\tau\kappa} \equiv \frac{1}{3!} {G_{\mu\nu\rho}}^{\alpha\beta\gamma}\epsilon_{\alpha\beta\gamma\sigma\tau\kappa} \, .
\end{split}
\end{equation}
 The above equation and symmetry ensure that $G$ transforms irreducibly in the $[0,0,4]$ of $\mathfrak{su}(4)$. 
It is easy to see that $[0,0,4]$ sits inside the tensor product of a pair of self-dual three-forms $F_3^+$ 
\begin{equation} \label{eq:2x2}
[0,0,2] \otimes [0,0,2] =[0,0,4]  \oplus [0,1,2]   \oplus [0,2,0]\ .
\end{equation}
For the $[0,1,2]$ part, we have
\begin{equation}
[0,1,0] \otimes [0,0,2] = [0,1,2] \oplus [1,0,1]\ .
\end{equation}
The representations $[0,2,0]$ and $[1,0,1]$ are immediately recognized  as the metric $g_{(\mu\nu)}$ and the two-form $B_{[\mu\nu]}$ (or $\Lambda^2 T^*\!\cM$) respectively. 
As before we build each individual $[0,0,2]$ by taking the tensor product of 2 positive chirality spinors
\begin{equation}
[0,0,1] \otimes [0,0,1] = [0,1,0] \oplus [0,0,2]\ .
\end{equation}
We consider tensor product of four chiral spinors $[0,0,1]$ and, applying the tensor product decomposition, obtain 
\begin{equation}
\begin{split}
[0,0,1]^{\otimes 4} & = \left( [0,1,0] \oplus [0,0,2] \right) \otimes \left( [0,1,0] \oplus [0,0,2] \right) \\
&= \left( [0,1,0] \otimes [0,1,0] \right) \oplus \left( [0,1,0] \otimes [0,0,2] \right) \oplus \left( [0,0,2] \otimes [0,1,0] \right) \\
& \;\;\;\;\oplus  [0,0,4]  \oplus [0,1,2]   \oplus [0,2,0]\ , 
\end{split}
\end{equation}
where  \eqref{eq:2x2} is used to get the last three terms.
The $[0,0,4]$ can now be extracted, and the result can be recast in terms of sections of corresponding bundles \cite{Minasian:2020vxn}
\begin{equation}
G \in \mathcal{C}^{\infty}\left(S^+\otimes [S^+ \otimes S^+  \otimes S^+ - (S^+ \otimes T^*\!\cM )^{\oplus 3} + (S^-)^{\oplus 2} ] \right)  + B + g\ .
\end{equation}
At this stage, we can state that the sections to which $B$ and $g$ belong do not contribute to the index density. Simply said, the metric and a generic two-form potential are anomaly-free. 

The relevant $(\mathbf 5,\mathbf 1)$ contribution to the index density is the factor
\begin{equation}
   \ch_{\bR_G}(\cR) = ( \ch_{+ \rm s}(\cR) )^3 -  3 \ch_{+\rm s}(\cR)  \ch_{\rm def}(\cR) + 2\ch_{- \rm s}(\cR) \,.
\end{equation}

\section{Anomalies}

\subsubsection*{$R$-Anomaly Cancellation}

Let us do a simpler computation of pure $R$-anomalies, first. Extracting the leading numbers 
from the gravitational contributions, we find for each fermionic chiral field
\bea
(\mathbf 2,\mathbf 1)&\rightarrow &\frac12 =\frac12\mathbb A(\cR)\biggr\vert_{\hbox{\scriptsize 0-form}}\ ,\cr\cr
(\mathbf 3,\mathbf 2)&\rightarrow &\frac{5}{2} =\frac12\mathbb A(\cR)\wedge (\ch_{\rm def}(\cR)-1)\biggr\vert_{\hbox{\scriptsize 0-form}}\ ,\cr\cr
(\mathbf 4,\mathbf 1)&\rightarrow &5 =\frac12 \mathbb A(\cR)\wedge (\ch_{\fat[2\fat]}(\cR)-   \ch_{\rm def}(\cR)+1)      \ ,     \biggr\vert_{\hbox{\scriptsize 0-form}}
\eea
while for the bosonic field $(\mathbf 3,\mathbf 1)$, represented by the
antisymmetric 2-form tensor,
\bea
(\mathbf 3,\mathbf 1)&\rightarrow &-2 =-\frac14\mathbb A(\cR)\wedge \ch_{\rm s}(\cR)\biggr\vert_{\hbox{\scriptsize 0-form}}\ ,
\eea
with the relative $-$ sign. The subscript $\fat[n\fat]$ means the $n$-th antisymmetric tensor of the
defining representation which is itself denoted by the subscript ``def."

For $d=6$ $\cN=(4,0)$, the pure $R$-anomaly is then,
\bea
P_8^{(4,0)}\biggr\vert_{\cR\rightarrow 0}&=&
\frac{1}{2}
\left( \ch_{\fat[3\fat]}(\cF_{\ksp(4)}) - 4\times \ch_{\fat[2\fat]}(\cF_{\ksp(4)}) + 10\times \ch_{\fat[1\fat]}(\cF_{\ksp(4)})\right)\biggr\vert_{\hbox{\scriptsize 8-form}} \cr\cr
&=&\frac{(\ii /2\pi)^4}{2\cdot 4!}
\left( {\rm tr}^{\ksp(4)}_{\fat[3\fat]}\cF^4 - 4\times {\rm tr}^{\ksp(4)}_{\fat[2\fat]}\cF^4 + 10\times {\rm tr}^{\ksp(4)}_{\fat[1\fat]=\rm def}\cF^4 \right)\ .
\eea
The relevant trace formulae were computed in appendix~\ref{app:A},
\bea
{\rm tr}^{\ksp(4)}_{\fat[2\fat]}\cF^4 &=& 3\times \left({\rm tr}^{\ksp(4)}_{\rm def}\cF^2 \right)^2 \cr\cr
{\rm tr}^{\ksp(4)}_{\fat[3\fat]}\cF^4 &=& -10\times {\rm tr}^{\ksp(4)}_{\rm def}\cF^4 +12\times \left({\rm tr}^{\ksp(4)}_{\rm def}\cF^2 \right)^2 \ ,
\eea
which shows immediately,
\bea
P_8^{(4,0)}\biggr\vert_{\cR\rightarrow 0}=0\ .
\eea
Hull's \cite{Hull:2000zn} proposal posits that $(4,0)$ tensor theory is actually an exotic realization of gravity where $(\mathbf 5, \mathbf 1; \mathbf 1)$ tensor has the metric
data hidden therein. If this proposal is valid, the gravitational and mixed
anomaly questions should be addressed with respect to the gauge symmetry of
this $(\mathbf 5, \mathbf 1; \mathbf 1)$ tensor field.

The pure $R$ part of the anomaly polynomial of $\cN=(3,1)$ tensor theory
goes similarly as
\bea
P_8^{(3,1)}\biggr\vert_{\cR\rightarrow 0}
&=&\frac{(\ii /2\pi)^4}{2\cdot 4!} \left(2\times  {\rm tr}_{\fat[2\fat]}(\cF_{\ksp(3)}^4) - {\rm tr}_{\fat[3\fat]}(\cF_{\ksp(3)}^4) -8\times {\rm tr}_{\fat[1\fat]}(\cF_{\ksp(3)}^4) + 5\times {\rm tr}_{\fat[1\fat]}(\cF_{\ksp(3)}^4)\right) \cr\cr
&&+\; \frac{(\ii /2\pi)^4}{2\cdot 4!}\left( 14\times {\rm tr}_{\fat[1\fat]}(\cF_{\ksp(1)}^4)-24\times {\rm tr}_{\fat[1\fat]}(\cF_{\ksp(1)}^4) +10\times{\rm tr}_{\fat[1\fat]}(\cF_{\ksp(1)}^4) \right)\cr\cr
&&+\; \frac{(\ii /2\pi)^4}{2\cdot 2!\cdot 2!} \left( {\rm tr}_{\fat[2\fat]}(\cF_{\ksp(3)}^2)\wedge  {\rm tr}_{\fat[1\fat]}(\cF_{\ksp(1)}^2)-4\times {\rm tr}_{\fat[1\fat]}(\cF_{\ksp(3)}^2)\wedge {\rm tr}_{\fat[1\fat]}(\cF_{\ksp(1)}^2) \right)\ ,
\eea
where we dropped the pure $\ksp(1)$ anomaly as their coefficients cancels away without
further manipulation. For the rest, we again read off the trace formulae from the previous chapter,
\bea
{\rm tr}^{\ksp(3)}_{\fat[2\fat]}\cF^2 &=& 4 \times {\rm tr}^{\ksp(3)}_{\rm def}\cF^2\ ,\cr\cr
{\rm tr}^{\ksp(3)}_{\fat[2\fat]}\cF^4 &=& -2 \times {\rm tr}^{\ksp(3)}_{\rm def}\cF^4 + 3\times \left({\rm tr}^{\ksp(3)}_{\rm def}\cF^2 \right)^2\ , \cr\cr
{\rm tr}^{\ksp(3)}_{\fat[3\fat]}\cF^4 &=& -7\times {\rm tr}^{\ksp(3)}_{\rm def}\cF^4 +6\times \left({\rm tr}^{\ksp(3)}_{\rm def}\cF^2 \right)^2\ .
\eea
The first of these three kills the mixed anomaly while the latter two shows
that pure $\ksp(3)$ anomaly vanishes as well,
\bea
P_8^{(3,1)}\biggr\vert_{\cR\rightarrow 0}=0\ .
\eea

\subsubsection*{Mixed-Anomaly Cancellation and the Gravitational Remainder}

Bolstered by the complete cancellation of the pure $R$-symmetry anomaly, let us
list the entire anomaly polynomial including the mixed and the gravitational. 
These latter have no particular reason to cancel away, a priori, yet, we find below
that the anomaly polynomial, when summed over the field content, leaves only
purely gravitational ones behind in both theories.

For the $\cN=(4,0)$ multiplet, the complete list is
\bea 
(\mathbf 5,\mathbf 1;\mathbf 1)&\rightarrow &-\frac12 \mathbb A(\cR) \left(  (\ch_{+ \rm s}(\cR))^3 - 3 \ch_{+ \rm s}(\cR) \ch_{\rm def}(\cR) + 2 \ch_{- \rm s}(\cR) \right)\biggr\vert_{\hbox{\scriptsize 8-form}}\ ,\cr\cr
(\mathbf 4,\mathbf 1;\mathbf 8)&\rightarrow &+\frac12 \mathbb A(\cR) \left(\ch_{\fat[2\fat]}(\cR)-   \ch_{\rm def}(\cR)+1\right)\ch_{\fat[1\fat]}(\cF_{\ksp(4)})  \biggr\vert_{\hbox{\scriptsize 8-form}}\ ,\cr\cr
(\mathbf 3,\mathbf 1;\mathbf{27})&\rightarrow & -\frac14\mathbb A(\cR)\ch_{\rm s}(\cR)\ch_{\fat[2\fat]}(\cF_{\ksp(4)}) \biggr\vert_{\hbox{\scriptsize 8-form}}\ ,\cr\cr
 (\mathbf 2,\mathbf 1;\mathbf{48})&\rightarrow &+\frac12\mathbb A(\cR)\ch_{\fat[3\fat]}(\cF_{\ksp(4)}) \biggr\vert_{\hbox{\scriptsize 8-form}}  \,.
\eea
We will write the gravitational part in terms of the Pontryagin classes $\mathbf{p}_n$ \cite{Nakahara:2003nw} 
and for the $R$-symmetry side use  the invariants $\hat{\mathbf q}_m$ as defined in \eqref{eq:gaugegroupinvariants},
modulo the necessary rescaling them by $1/{(2\pi)^{2m}}$ into $\mathbf q_m$. 

We already saw in the previous section that the pure $R$-symmetry part, i.e., those involving ${\mathbf q}_m$'s only cancel out. The rest with  the gravitational $\mathbf{p}_n$ part present are 
\bea
(\mathbf 5,\mathbf 1;\mathbf 1)&\rightarrow &-\frac12 \times \frac{1}{5760}  (3840 \mathbf{p}_1^2 + 19200 \mathbf{p}_2) \ ,\cr\cr
(\mathbf 4,\mathbf 1;\mathbf 8)&\rightarrow &+\frac12 \times \frac{1}{5760} ( 6320 \mathbf{p}_1^2 + 22720 \mathbf{p}_2 - 7440 \mathbf{p}_1 \mathbf{q}_1 + \ldots)\ , \cr\cr 
(\mathbf 3,\mathbf 1;\mathbf{27})&\rightarrow & -\frac14 \times \frac{1}{5760} (-1728 \mathbf{p}_1^2 + 12096 \mathbf{p}_2 - 11520 \mathbf{p}_1 \mathbf{q}_1 + \ldots) \ ,\cr\cr 
 (\mathbf 2,\mathbf 1;\mathbf{48})&\rightarrow &+\frac12\times \frac{1}{5760}(336 \mathbf{p}_1^2 - 192 \mathbf{p}_2 + 1680 \mathbf{p}_1 \mathbf{q}_1 + \ldots)\,, 
\eea
where the ellipses refer to pure $R$ parts. These together give
\bea
P_8^{(4,0)}= \frac{1}{72} (23 \mathbf{p}_1^2 - 17 \mathbf{p}_2) \neq 0 \ ,
\eea
leaving behind pure gravitational anomalies only. 

The computation for the $\cN=(3,1)$ multiplet is analogous. The index density contribution are listed as
\bea
(\mathbf 4, \mathbf 2; \mathbf 1, \mathbf 1)&\rightarrow & -\frac12 \mathbb A(\cR) \left( ( \ch_{+ \rm s}(\cR) )^2 \ch_{- \rm s}(\cR)-  2 \ch_{- \rm s}(\cR)  \ch_{\rm def}(\cR) - 2\ch_{+ \rm s}(\cR)\right)\biggr\vert_{\hbox{\scriptsize 8-form}}\ ,\cr\cr
(\mathbf 4, \mathbf 1; \mathbf 1, \mathbf 2)&\rightarrow &+\frac12 \mathbb A(\cR) \left(\ch_{\fat[2\fat]}(\cR) -   \ch_{\rm def}(\cR)+1\right)\ch_{\fat[1\fat]}(\cF_{\ksp(1)}) \biggr\vert_{\hbox{\scriptsize 8-form}}\ ,\cr\cr
(\mathbf 3, \mathbf 2; \mathbf 6, \mathbf 1)&\rightarrow & + \frac12 \mathbb A(\cR)(\ch_{\rm def}(\cR)-1)\ch_{\fat[1\fat]}(\cF_{\ksp(3)}) \biggr\vert_{\hbox{\scriptsize 8-form}}\ ,\cr\cr
(\mathbf 3,\mathbf 1;\mathbf 6, \mathbf 2)&\rightarrow & -\frac14\mathbb A(\cR)\ch_{\rm s}(\cR)\ch_{\fat[1\fat]}(\cF_{\ksp(3)})\ch_{\fat[1\fat]}(\cF_{\ksp(1)}) \biggr\vert_{\hbox{\scriptsize 8-form}}\ ,\cr\cr
 (\mathbf 2, \mathbf 1; \mathbf{14}, \mathbf 2)&\rightarrow &+\frac12\mathbb A(\cR)\ch_{\fat[2\fat]}(\cF_{\ksp(3)})\ch_{\fat[1\fat]}(\cF_{\ksp(1)}) \biggr\vert_{\hbox{\scriptsize 8-form}}\ ,\cr\cr
 (\mathbf 1, \mathbf 2; \mathbf{14}', \mathbf 1)&\rightarrow &-\frac12\mathbb A(\cR)\ch_{\fat[3\fat]}(\cF_{\ksp(3)}) \biggr\vert_{\hbox{\scriptsize 8-form}} \, .
\eea
{}From this, we obtain the anomaly polynomials, again with pure $R$ part suppressed as ellipses,
given the eventual cancellation above,
\bea
(\mathbf 4, \mathbf 2; \mathbf 1, \mathbf 1)&\rightarrow & -\frac12 \times \frac{1}{5760}  ( 7616 \mathbf{p}_1^2 + 15808 \mathbf{p}_2)\ , \cr\cr
(\mathbf 4, \mathbf 1; \mathbf 1, \mathbf 2)&\rightarrow &+\frac12 \times \frac{1}{5760}  ( 1580 \mathbf{p}_1^2 + 5680 \mathbf{p}_2 - 7440 \mathbf{p}_1 \mathbf{\tilde{q}}_1 + \ldots) \ ,\cr\cr 
(\mathbf 3, \mathbf 2; \mathbf 6, \mathbf 1)&\rightarrow & + \frac12 \times \frac{1}{5760}  ( 1650 \mathbf{p}_1^2 - 5880 \mathbf{p}_2 - 2280\mathbf{p}_1 \mathbf{q}_1 + \ldots) \ ,\cr\cr 
(\mathbf 3,\mathbf 1;\mathbf 6, \mathbf 2)&\rightarrow & -\frac14\times \frac{1}{5760}  ( -768 \mathbf{p}_1^2 + 5376 \mathbf{p}_2 - 3840 \mathbf{p}_1 \mathbf{q}_1 - 11520 \mathbf{p}_1 \mathbf{\tilde{q}}_1 +  \ldots ) \ ,\cr\cr 
 (\mathbf 2, \mathbf 1; \mathbf{14}, \mathbf 2)&\rightarrow &+\frac12\times \frac{1}{5760}  ( 196 \mathbf{p}_1^2 - 112 \mathbf{p}_2 + 960 \mathbf{p}_1 \mathbf{q}_1  +1680 \mathbf{p}_1 \mathbf{\tilde{q}}_1 + \ldots)\ , \cr\cr 
 (\mathbf 1, \mathbf 2; \mathbf{14}', \mathbf 1)&\rightarrow &-\frac12\times \frac{1}{5760}  ( 98 \mathbf{p}_1^2 - 56 \mathbf{p}_2 + 600 \mathbf{p}_1 \mathbf{q}_1 + \ldots) \, ,
\eea
where $\mathbf q_m$'s are $\ksp(3)$-curvatures invariants while $\mathbf{\tilde{q}}_m$'s stand for $\ksp(1)$ invariants.
In the end we find the total 
\bea
P_8^{(3,1)}= \frac{1}{180} (-61 \mathbf{p}_1^2 - 293 \mathbf{p}_2) \neq 0 \,.
\eea
which is again purely gravitational.

In conclusion, all anomalies associated with $R$-symmetries, {\it including the mixed ones, }
cancel out completely, leaving behind a net gravitational anomaly. The latter does not 
come in a factorized  form, so a counter-term of the Green-Schwarz type is not possible.

\section{Further Thoughts}

We found a common and mysterious behavior of perturbative anomalies 
of the two maximally supersymmetric tensor theories in $d=6$; purely gravitational anomalies
do not cancel out, yet the pure $R$-symmetry anomalies and the mixed anomalies involving 
the diffeomorphism and the $R$-symmetry cancel out entirely. The left-over gravitational anomaly is no 
mystery, given how the supermultiplet does not contain the metric; we have no right to
expect the general covariance to hold even classically, not to mention at the quantum level,
in the absence of the metric as a dynamical field. Therefore,  the anomalous diffeomorphism 
is no big deal. The role of the general covariance is replaced by various $n$-form gauge symmetries that 
shifts the chiral tensor fields.

What does beg for questions, perhaps with significant implications hidden, is 
why the anomalies involving $R$-symmetry, in particular the mixed ones, all cancel out. 
Let us sit back and recall what happens with other maximal  supergravities. 
Typically the supermultiplet includes a metric, so that the entire superalgebra is gauged.
Gauging of $R$-symmetries is a little more subtle since the associated gauge fields are
often frozen on-shell and do not offer dynamical gauge fields. The gauging 
can be seen directly only if we go sufficiently off-shell. Nevertheless,
the gauging of the $R$-symmetries is unavoidable since the superalgebra
starts from the translation invariance which is gauged by coupling to metric,
and an incomplete gauging of the superalgebra would be inconsistent with the superalgebra itself. 

This does not necessarily mean the one-loop anomalies must cancel out since the anomaly inflow mechanism
is in principle possible. For models with less than maximal supersymmetries, such inflows are
generically needed and realized in string theory embeddings. However, for maximal supergravities, 
there is often no obvious inflow mechanism from the string theory side, except for the symmetries 
that involve $U(1)_{R}$ to be canceled by a Green-Schwarz  mechanism. Instead, the maximal 
supergravities tend to show complete cancellation of the one-loop anomalies of diffeomorphism 
and non-Abelian $R$-symmetries. 

To understand the anomaly structure we saw for the maximally supersymmetric 
tensor theories, we need some inkling on how bosonic part of the superalgebra might be
gauged. Regardless of how this is achieved,  one expects that at least the 
$R$-symmetry would be eventually gauged in the conventional manner, for which the pure 
$R$-symmetry anomaly cancellation is certainly good news. It is the cancellation of the
diffeomorphism and $R$-symmetry mixed anomalies that is puzzling. 

Chris Hull, who suggested how these theories reduce to $d=5$ supergravity upon a dimensional reduction,
has also suggested \cite{Hull:2000rr} that these two chiral superalgebras may admit a modified version where 
the diffeomorphism would be replaced by exotic gauge transformations that shift 
the bosonic chiral tensors similar to how gauge symmetries act on 1-form gauge connections. These are analogous
to how a massless 2-form $B$ admits 1-form gauge symmetry. 

With Hull's proposal, the would-be diffeomorphisms of 
$d=5$ can be seen to elevate to $d=6$ higher-form gauge transformations, so that one must demand a quantum Ward 
identities for these generalized forms of ``diffeomorphism,"  
rendering the above nonvanishing pure gravitational anomaly aside as irrelevant. 
In this viewpoint, the cancellation of the mixed anomaly we have
observed become quite mysterious, not that it conflicts with these thoughts.

Eventually, we need to understand how to quantify anomalies associated with this generalized notion of ``diffeomorphisms". Although anomalies associated with higher-form
symmetries are now discussed in various literature (see for example the landmark article \cite{Gaiotto:2014kfa}), they are often in the form of discrete
versions. For instance, the object that would minimally couple to a continuous 2-form 
gauge field $B$ are strings, so the familiar anomaly computation that starts from
the path integral of chiral quantum fields are no longer useful. Nor are such relevant
for the problem at hand, since the tensor theory would consist of local fields only, however
unusual they might be. 

We must have a new computational scheme where chiral bosons and chiral fermions of 
these tensor multiplets are coupled in some definite manner, and  where we can perform
Feynman diagram computation or a heat-kernel computation for the anomalous phase. 
If we take the suggestion by Hull seriously, there must be a definite set of 
couplings in $d=6$ as implied by $d=5$ maximal supergravity, which would be the
right starting point for the relevant anomaly discussion. We leave this challenge 
to a future endeavor.

\section*{Acknowledgments}
We thank Qiang Jia, Victor Lekeu, Hong Lü, Ruben Minasian, Yi Pang and Yi-Nan Wang for discussions. PY is particularly grateful to Sungjay Lee for bringing the $(4,0)$ theory and Hull's proposal to his attention and also for several key discussions thereafter. 
YZ would like to thank Korea Institute for Advanced Study for their warm hospitality during this project. We are
also grateful to the 18th Kavli Asian Winter School on Strings, Particles and Cosmology with the meeting number: YITP-W-23-13. 
YZ is supported by National Science Foundation of China under Grant No. 12305077 and under Grant No. 12175004 and also 
by the Office of China Postdoc Council (OCPC) and Peking University under Grant No. YJ20220018. PY is supported by KIAS individual grant (PG005705).

\appendix

\section{Batalin–Vilkovisky (BV) field-antifield formalism}
\label{app:BV}
The Batalin–Vilkovisky (BV) field-antifield formalism \cite{Batalin:1981jr,Batalin:1984jr} provides a systematic way of quantizing gauge systems that are in general very involved. Due to our interests and purpose, we give a compact review of only the BV-quantisation algorithm for first-stage reducible theories, and the notation and conventions mostly follow the comprehensive reviews \cite{Gomis:1994he,Henneaux:1992ig}.
A detailed and complete discussion on applying this formalism to the exotic gravitino $\psi_{\mu\nu}$ can be found in \cite{Lekeu:2021oti}.
\paragraph{Gauge structure of first-stage reducible system.} Let $\varphi^i$ denote the dynamical field variables in our theory. Our starting point is an action $S_0[\varphi^i]$ which is invariant under $m$ gauge invariances $\delta \varphi^i = R^i_\alpha \Lambda^\alpha$ with generator $R^i_\alpha$ and parameter $\Lambda^\alpha$.
Furthermore, we demand that the above gauge transformations to be invariant under $n$ (reducibility) `gauge-for-gauge' transformations $\delta \Lambda^\alpha = Z^\alpha_a \lambda^a$, and we assume that there are \emph{no} further reducibilities. 

In other words, this reducible gauge invariance is equivalent to the following statements
\begin{equation}
    \frac{\delta^R S_0}{\delta \varphi^i} R^i_\alpha = 0 \, , \quad R^i_\alpha Z^\alpha_a = 0 \, , \quad Z^\alpha_a \; \text{are linearly independent on-shell} \, ,
\end{equation}
with $\alpha = 1, \dots, m$ and $a = 1, \dots, n$.
Here a contracted index includes space-time integration and the superscript $R$ (resp.~$L$) indicates that the derivative is acting from the right (resp. left)\footnote{For any function or functional $X$ of the field $\phi$ we have the variation $\delta X(\phi)=\delta \phi \frac{\delta^{L} X}{\delta \phi}=\frac{\delta^{R} X}{\delta \phi} \delta \phi$.}. 

In such set-up, the gauge-for-gauge transformations terminate at the first step and are hence called \emph{first-stage reducible} theories. The number of independent gauge redundancies in the fields $\varphi^i$ is therefore equal $m-n$. We can fix these gauge redundancies by imposing an appropriate gauge-fixing condition $\chi^\alpha(\varphi) = 0$ that satisfies $n$ constraints:
\begin{equation}\label{eq:constrainedchi}
X_{a\alpha} \,\chi^\alpha(\varphi) = 0
\end{equation}
with $X_{a\alpha}$ of maximal rank. 
\paragraph{BV-minimal sector.}
To perform the BV-quantisation, we first introduce the ghost $C^\alpha$ corresponding to the gauge parameter $\Lambda^\alpha$, and the ghost-for-ghost $c^a$ corresponding to the reducibility parameter $\lambda^a$.
\emph{Grassmann parity} will be denoted as $\epsilon(\varphi_i) \equiv \epsilon_i$, $\epsilon(\Lambda^\alpha) \equiv \epsilon_\alpha$ and etc, and as for the ghosts, they are always parity-opposed to their gauge parameters: $\epsilon(C^\alpha) = \epsilon(\Lambda^{\alpha}) + 1$ and $\epsilon(c^a) = \epsilon(\lambda^a) + 1$. We have ghosts from two different stages, to distinguish between them, we introduce the \emph{ghost numbers} denoted as $\gh(X)$ for a field $X$. For our classical fields, $\gh(\varphi^i) = 0$ whereas for the ghosts they are just the stage number plus one, i.e. $\gh(C^\alpha) = 1$ and $\gh(c^a) = 2$.

We then double the space of field variables by assigning to each field $\Phi^I$ (collectively notation for the set of all fields) its \emph{antifields} $\Phi^*_I$.
The ghost number assignments and Grassmann parities of the antifields are related to their ordinary counterparts
\begin{equation}
   \gh(\Phi^*_I) = - \gh(\Phi^I) - 1 \, ,\quad \epsilon(\Phi^*_I) = \epsilon(\Phi^I) + 1 \quad (\text{mod}\,2)\, .
\end{equation}
The set $\{\varphi^i, C^\alpha, c^a; \varphi^*_i, C^*_\alpha, c^*_a\}$ is called the \emph{minimal} sets of fields and antifields. 

The action $S_0[\varphi^i]$ is then extended to the minimal BV action $\SM[\varphi^i, C^\alpha, c^a; \varphi^*_i, C^*_\alpha, c^*_a]$
depending on the original fields $\varphi^i$ but also on the ghost fields $C^\alpha$, $c^a$ and their antifields $\varphi^*_i$, $C^*_\alpha$ and $c^*_a$. It is a ghost number zero, even functional that should be a proper solution of the \emph{classical master equation}
\begin{equation}
(\SM, \SM) = 0\, ,
\end{equation}
where the \emph{antibracket} $(\, \cdot \, , \, \cdot \,)$ is defined as
\begin{equation}
    (X,Y) = \frac{\delta^R X}{\delta \Phi^I}\frac{\delta^L Y}{\delta \Phi^*_I} - \frac{\delta^R X}{\delta \Phi^*_I}\frac{\delta^L Y}{\delta \Phi^I}\, .
\end{equation}
Moreover, it should satisfy the boundary condition:
\begin{equation}
S_0[\varphi^i] = \SM[\Phi, \Phi^* = 0]\, .
\end{equation}
These two conditions completely determine $\SM$, which always exists; it starts with
\begin{equation}
\SM[\varphi^i, C^\alpha, c^a; \varphi^*_i, C^*_\alpha, c^*_a] = S_0[\varphi] + \varphi^*_i R^i_\alpha C^\alpha + C^*_\alpha Z^\alpha_a c^a + \dots
\end{equation}
and the omitted terms carry explicit information about the gauge algebra, on-shell closure, etc.
\paragraph{Gauge-fixing and non-minimal sector.} We introduce \emph{three} extra trivial pairs: $(C'^\alpha, b^\alpha)$ to fix the gauge freedom of $\varphi^i$, but also two more, $(c'^a, \pi^a)$ and $(\eta^a, \pi'^a)$, to fix the gauge freedom of the ghosts $C^\alpha$ and $C'^\alpha$ themselves. This is depicted in figure \ref{fig:pairsfirststage}. Their ghost numbers and Grassmann parities can be found in table \ref{tab:reducible}.

The \emph{non-minimal} action is then
\begin{equation}
    \SNM = \SM + C'^*_\alpha b^\alpha + c'^*_a \pi^a + \eta^*_a \pi'^a\,,
\end{equation}
which still satisfies the master equation. \\
\begin{figure}
\centering
\begin{tikzpicture}
\node (ph) at (0,0) {$\varphi^i$};
\node (C0) at (1,-1) {$C^{\alpha}$};
\node (C1) at (2,-2) {$c^{a}$};
\node (C0p) at (-1,-1) {${C'}^{\alpha}$};
\node (C1p) at (-2,-2) {$\eta^{a}$};
\node (e) at (0,-2) {${c'}^{a}$};
\foreach \from/\to in {ph/C0, C0/C1}
	\draw [very thick] (\from) -- (\to);
\foreach \from/\to in {ph/C0p, C0p/C1p, C0/e}
	\draw [->] (\from) -- (\to);
\end{tikzpicture}
\qquad\qquad
\begin{tikzpicture}
\node (b) at (0,0) {$b^{\alpha}$};
\node (p1) at (1,-1) {$\pi^{a}$};
\node (p1p) at (-1,-1) {$\pi'^{a}$};
\foreach \from/\to in {b/p1, b/p1p}
	\draw [dashed] (\from) -- (\to);
\end{tikzpicture}
\caption{The pyramid of ghosts fields in the first-stage reducible case. The fields linked by a thick line constitute the minimal BV sector; an arrow $a \rightarrow b$ indicates that the field $b$ (along with its partner in a trivial pair) is introduced to fix the gauge freedom of $a$. The second pyramid shows the partners of the non-minimal fields of the first pyramid.}
\label{fig:pairsfirststage}
\end{figure}
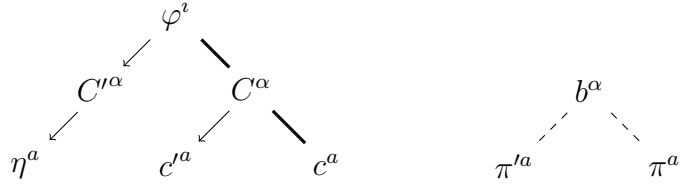
\begin{table}
    \centering
    \begin{tabular}{c| c c c c c c | c c c}
         & $\varphi^i$ & $C^\alpha$ & $C'^\alpha$ & $c^a$ & $c'^a$ & $\eta^a$ & $b^\alpha$ & $\pi^a$ & $\pi'^a$ \\ \midrule
        $\gh$ & $0$ & $1$ & $-1$ & $2$ & $-2$ & $0$ & $0$ & $-1$ & $1$ \\[\defaultaddspace]
        $\epsilon$ & $\epsilon_i$ & $\epsilon_\alpha+1$ & $\epsilon_\alpha+1$ & $\epsilon_a$ & $\epsilon_a$ & $\epsilon_a$ & $\epsilon_\alpha$ & $\epsilon_a+1$ & $\epsilon_a+1$
    \end{tabular}
    \caption{Ghost numbers and Grassmann parities of the various fields in the first-stage reducible case. Antifields have $\gh(\Phi^*_I) = - \gh(\Phi^I) - 1$ and opposite parity, $\epsilon(\Phi^*_I) = \epsilon(\Phi^I) + 1$.}
    \label{tab:reducible}
\end{table}
To eliminate the antifields, we consider the surfaces in the phase space given by the following condition\footnote{There is no need to distinguish between left or right derivatives of $\Psi$ with respect to the fields $\frac{\delta \Psi(\Phi)}{\delta \Phi^A} \equiv \frac{\delta^L \Psi(\Phi)}{\delta \Phi^A} = (-1)^{\epsilon_A(\epsilon(\Psi)+1)} \frac{\delta^R \Psi(\Phi)}{\delta \Phi^A} =  \frac{\delta^R \Psi(\Phi)}{\delta \Phi^A}$.}
\begin{equation}
    \Sigma = \bigg\{ \{\Phi, \Phi^*\} |  \Phi^*_A  = \frac{\delta \Psi(\Phi)}{\delta \Phi^A}  \bigg\} \, ,
\end{equation}
where $\Psi$, called the \emph{gauge-fixing fermion}, is a Grassmann odd functional with ghost number $-1$, i.e.
\begin{equation}
    \epsilon(\Psi) = 1 \, , \quad \gh(\Psi) = -1 \, .
\end{equation}
When a functional $X$ is restricted on the surface $\Sigma$, we mean to replace the dependence of $X$ on the antifields $\Phi^*_A$ simply by $\frac{\delta \Psi(\Phi)}{\delta \Phi^A}$.
The tricky point here is to find an \emph{admissible} $\Psi$ with which correct degrees of freedom in $\varphi^i$ are fixed, and the existence of propagators is ensured when the action is expanded about a solution of the equations of motion. In fact, one can show \cite{Gomis:1994he} that the path integral quantization is independent of the choice of admissible gauge-fixing fermions.
We discuss here two choices of $\Psi$ for the first-stage reducible theories.
\begin{itemize}
    \item \emph{Delta-function gauge-fixing.} We simply take the well-known \cite{Gomis:1994he} gauge-fixing fermion:
\begin{equation}
    \Psi_\delta = C'_\alpha \chi^\alpha(\varphi) + c'^a \omega_{a\alpha} C^\alpha + \eta^a \sigma_{a}^{\alpha} C'_\alpha\, ,
\end{equation}
where $\omega$ and $\sigma$ are of maximal rank and we take the pair $(C'_\alpha, b_\alpha)$ to have indices down for this paragraph only. The gauge-fixed action then reads
\begin{align}
    S_\delta &\equiv \SNM\left[ \Phi^I, \Phi^*_I = \frac{\delta \Psi_\delta}{\delta \Phi^I} \right] \\
    &= \SM\left[ \varphi^i, C^\alpha, c^a; \varphi^*_i = C'_\alpha \frac{\delta^R \chi^\alpha}{\delta \varphi^i}, C^*_\alpha = c'^a \omega_{a\alpha}, c^*_a = 0 \right] \nn\\
    &\qquad + (\chi^\alpha(\varphi) + \eta^a \sigma_{a}^{\alpha})\, b_\alpha + \omega_{a\alpha} C^\alpha \pi^a + \sigma_{a}^{\alpha} C'_\alpha \pi'^a \, . \label{eq:reddelta}
\end{align}
As we can see, the ghosts $C^\alpha$ and $C'_\alpha$ are both gauge fields. Their gauge invariances are fixed by the $2 m$ gauge conditions $\omega_{a\alpha} C^\alpha = 0$ and $\sigma_{a}^{\alpha} C'^\alpha = 0$ imposed by the auxiliary fields $\pi^a$ and $\pi'^a$. The field $b^\alpha$ is also auxiliary and imposes the equation
\begin{equation}\label{eq:chieta}
    \chi^\alpha(\varphi) + \eta^a \sigma_{a}^{\alpha} = 0\, .
\end{equation}
Among these $m$ conditions, $m-n$ fix the gauge invariance of the original fields $\varphi^i$, and the remaining $n$ set the extra ghost $\eta$ to zero.

\item \emph{Gaussian gauge-fixing.} Alternatively, one can also get the gauge-fixing term $\cD_{\alpha\beta} \chi^\beta \chi^\alpha$. The crucial ingredient is to perform a \emph{canonical transformation}, which was first discovered for irreducible theories \cite{Batalin:1983ar} and can be applied to the first-stage reducible systems as well \cite{Lekeu:2021oti}:
\begin{align}
    b^\alpha &\rightarrow \tilde{b}^\alpha = b^\alpha - \chi^\alpha(\varphi) \nn\\
    \varphi^*_i &\rightarrow \tilde{\varphi}^*_i = \varphi^*_i + b^*_\alpha \frac{\delta^R \chi^\alpha}{\delta \varphi^i}\label{eq:canonical}
\end{align}
with other fields unchanged. We then take the gauge-fixing fermion
\begin{equation}
\Psi_\text{G} = \frac{1}{2} C'^\alpha \cD_{\alpha\beta}(\varphi) \left( \chi^\beta(\varphi) + b^\beta \right) + c'^a \omega_{a\alpha} C^\alpha + \eta^a \sigma_{a\alpha} C'^\alpha\, ,
\end{equation}
which is of the same form as $\Psi_\delta$, with only the first term modified. Eliminating the antifields using $\Psi_G$ then gives
\begin{align}
    S_\text{G} &\equiv \tSNM\left[ \Phi^I, \Phi^*_I = \frac{\delta \Psi_\text{G}}{\delta \Phi^I} \right] \\
    &= \SM\left[ \varphi^i, C^\alpha, c^a; C'^\alpha \cD_{\alpha\beta} \frac{\delta^R \chi^\beta}{\delta \varphi^i} + \frac{1}{2} C'^\alpha \frac{\delta^R \cD_{\alpha\beta}}{\delta \varphi^i}\left( \chi^\beta + b^\beta \right) (-1)^{\epsilon_i \epsilon_\beta}, c'^a \omega_{\alpha a} ,  0 \right] \nonumber \\
    &\quad-\frac{1}{2} \cD_{\alpha\beta} \chi^\beta \chi^\alpha + \frac{1}{2}\cD_{\alpha\beta} b^\beta b^\alpha \\
    &\quad+ \eta^a \sigma_{a\alpha} (b^\alpha - \chi^\alpha) + \pi^a \omega_{a \alpha} C^\alpha  + \pi'^a \sigma_{a\alpha} C'^\alpha \, . \nonumber
\end{align}
Because of the constraint \eqref{eq:constrainedchi} satisfied by $\chi^\alpha(\varphi)$, there is a privileged choice for the matrix $\sigma_{a\alpha}$: simply take $\sigma = X$. This gets rid of the unwanted term $\eta^a \sigma_{a\alpha} \chi^\alpha$ in the last line, and one remains with
\begin{align}
    S_\text{G} = \SM&\left[ \varphi^i, C^\alpha, c^a; C'^\alpha \cD_{\alpha\beta} \frac{\delta^R \chi^\beta}{\delta \varphi^i} + \frac{1}{2} C'^\alpha \frac{\delta^R \cD_{\alpha\beta}}{\delta \varphi^i}\left( \chi^\beta + b^\beta \right) (-1)^{\epsilon_i \epsilon_\beta}, c'^a \omega_{\alpha a} ,  0 \right] \nn \\
    &-\frac{1}{2} \cD_{\alpha\beta} \chi^\beta \chi^\alpha + \frac{1}{2}\cD_{\alpha\beta} b^\beta b^\alpha + \pi^a \omega_{a \alpha} C^\alpha  + \pi'^a X_{a\alpha} C'^\alpha + \eta^a X_{a\alpha} b^\alpha\, , \label{eq:gaussianfirststage}
\end{align}
featuring the desired gauge-breaking term $\cD_{\alpha\beta} \chi^\beta \chi^\alpha$. We see that the field $b^\alpha$ is propagating whenever $\cD$ contains derivatives, and couples to the other fields and ghosts if $\cD$ is field-dependent. This generalizes a result of the aforementioned irreducible theories (e.g. Rarita-Schwinger field) \cite{Batalin:1983ar} to the first-stage reducible case.

Integrating over the auxiliary fields $\pi^a$ and $\pi'^a$ will impose the gauge conditions
\begin{equation}
\omega_{a \alpha} C^\alpha = 0\, , \quad X_{a\alpha} C'^\alpha = 0
\end{equation}
on the ghost fields $C^\alpha$ and $C'^\alpha$, as in the delta-function gauge-fixing case. On the other hand, $\eta^a$ plays here a very different role as it did in \eqref{eq:reddelta}: it is now a Lagrange multiplier for the constraint
\begin{equation}
X_{a\alpha} \, b^\alpha = 0
\end{equation}
on the field $b^\alpha$. Notice how both $C'^\alpha$ and $b^\alpha$ satisfy the same constraint as $\chi^\alpha(\varphi)$ in this gauge-fixing scheme.

\end{itemize}

\section{Anomaly Polynomials for $SO$ and $Sp$}
\label{app:A}

Let us make a quick overview of relations among symmetric traces between
various representations.\footnote{The content of this Appenidx is borrowed from a textbook in writing by the senior author\cite{GQFT}.}
For this, it suffices to consider $\cF$ of the form
\bea
\ii\cF\qquad \rightarrow \qquad \mathfrak f_IH^I
\eea
with the Cartan generators $H$'s.  To relate the anomaly polynomials 
in other representations to those of the defining ones,
\bea \label{eq:gaugegroupinvariants}
\hat{\mathbf q}_m(\mathfrak f)\equiv \sum_i 2 (\mathfrak f_i)^{2m}
\eea
will be used as the convenient unit expressions in the following.
One universal fact to keep in mind is that
\bea
{\rm tr}_{\mathbf R}(\mathfrak f_IH^I)^{2m+1} =0
\eea
for any (pseudo-)real representation $\mathbf R$. For $\kso$ and
$\ksp$, this holds for the defining representations and all tensor
products thereof.

For a quick summary of notations, $\mathbb{A}$ is the A-roof genus, ${\rm ch}_\bR$ is
the Chern character via the trace in the representation $\bR$. The notations for representation ``adj" and ``def"
should be self-explanatory, while $\fat[n\fat]$ is the $n$-antisymmetric tensor of ``def." Spinors
are denoted by ``s" while (anti-)chiral spinors are labeled as ``$\pm$s." The trace computations
are performed by summing over the weights $\mu$ that constitute the representation $\bR$ in question,
\bea
{\rm tr}_{\mathbf R}(\,\cdots) =\sum_{\vert\mu\rangle \in \bR} \langle \mu\vert (\,\cdots)\vert \mu\rangle \ .
\eea
The normalization choice below is such that short roots of $\kso(2r+1)$ are $\pm \ke_i$'s with the unit length
while the long roots of $\ksp(r)$ are $\pm 2\ke_i$'s.

\subsection{$\kso(N)$ }

Let us start with the simplest case of $\kso(N=2r)$.
\bea
{\rm tr}_{\rm adj}^{\kso(2r)}(\mathfrak f_IH^I)^{2m} &=&4(r-1)\times \sum_{i=1}^r (\mathfrak f_i)^{2m} \cr\cr
&&+4\times \sum_{i<j} \sum_{l=1}^{m-1} \frac{(2m)!}{(2l)!(2m-2l)!}
(\mathfrak f_i)^{2l}(\mathfrak f_j)^{2(m-l)}
\eea
where $4(r-1)$ factor for $\kf_i^{2m}$ is from  summing over $j\neq i$ the contributions
from the roots $\pm\ke_i\pm\ke_j$. This results in
\bea
{\rm tr}_{\rm adj}^{\kso(2r)}(\mathfrak f_IH^I)^{2}& =& 
(2r-2) \times \hat{\mathbf q}_1(\mathfrak f)\cr\cr
{\rm tr}_{\rm adj}^{\kso(2r)}(\mathfrak f_IH^I)^{4} 
&=& (2r-8)\times  \hat{\mathbf q}_2(\mathfrak f)  + 3\times  \hat{\mathbf q}_1(\mathfrak f)^2\cr\cr
{\rm tr}_{\rm adj}^{\kso(2r)}(\mathfrak f_IH^I)^{6} 
&=& (2r-32)\times  \hat{\mathbf q}_3(\mathfrak f)  + 15\times  \hat{\mathbf q}_1(\mathfrak f)\, \hat{\mathbf q}_2(\mathfrak f)
\eea
and so on.

One can easily extend this to $\kso(2r+1)$. Since the root system is extended to include $\pm\ke_i$'s
in addition to those of $\kso(2r)$, this has the effect of adding a contribution of the defining
representation,
\bea
{\rm tr}_{\rm adj}^{\kso(2r+1)}(\mathfrak f_IH^I)^{2m} &=&(4(r-1)+2)\times \sum_{i=1}^r (\mathfrak f_i)^{2m} +\cdots 
\eea
which, relative to the above $\kso(2r)$, merely shift the coefficient of the highest 
power summation $\hat{\mathbf q}_{m}(\mathfrak f)$. The two sets of results combine naturally to
\bea
{\rm tr}_{\rm adj}^{\kso(N)}\cF^{2} &=&(N-2) \times {\rm tr}_{\rm def}^{\kso(N)}\cF^{2} \\\cr
{\rm tr}_{\rm adj}^{\kso(N)}\cF^{4} &=&(N-8) \times {\rm tr}_{\rm def}^{\kso(N)}\cF^{4} + 3\times \left( {\rm tr}_{\rm def}^{\kso(N)}\cF^{2}\right)^2\cr\cr
{\rm tr}_{\rm adj}^{\kso(N)}\cF^{6} &=&(N-32) \times {\rm tr}_{\rm def}^{\kso(N)}\cF^{6} + 15\times \left( {\rm tr}_{\rm def}^{\kso(N)}\cF^{2}\right)\left( {\rm tr}_{\rm def}^{\kso(N)}\cF^{4}\right)\nonumber
\eea
and so on.

\subsubsection*{Spinors}

Spinor representations of $\kso(N)$ deserve special attention.
The trace formulae for spinor can be written most universally for
both $\kso(2r+1)$ and  $\kso(2r)$ via the Chern character,
\bea
\ch_{\rm s}^{\kso(N)}(\cF)&=&{\rm tr}_{\rm s} e^{\ii\cF/2\pi}\; =\;
\sum_{\mu^{\rm s}}\langle \mu^{\rm s}\vert  e^{\sum_I\kf_IH^I/2\pi}\vert \mu^{\rm s}\rangle\cr\cr
&=& \prod_{i=1}^r\sum_\pm \biggr\langle \pm \frac12\ke_i\biggr\vert  e^{\kf_i H^i/2\pi}\biggr\vert \pm \frac12\ke_i \biggr\rangle\cr\cr
&=&\prod_{i=1}^r 2\cosh(\kf_i/4\pi)
\eea
with the subscript/superscript s for spinors. 

For $\kso(2r)$, there is another invariant to compute since the spinor
splits into two irreducible representations, $\pm$s for chiral/anti-chiral spinors. 
Instead of computing them individually, we will compute the difference
\bea
\ch_{+\rm s}^{\kso(2r)}(\cF)-\ch_{-\rm s}^{\kso(2r)}(\cF)&=&{\rm tr}_{+\rm s}^{\kso(2r)} e^{\ii\cF/2\pi}-{\rm tr}_{-\rm s}^{\kso(2r)} e^{\ii\cF/2\pi}\cr\cr
&=&\prod_{i=1}^r 2\sinh(\kf_i/4\pi)\cr\cr
&=& \frac{\prod_i (\kf_i/2\pi)}{\mathbb A(\cF)} =\frac{\chi(\cF)}{\mathbb A(\cF)}
\eea
where the numerator, the Pfaffian of $\cF/2\pi$ in the defining representation,
is the Euler invariant while we see the A-roof genus downstairs.

To relate the results to those from the defining representation,
we may expand the above, modulo the factors of $2\pi$'s, as follows.
With $\kso(2r+1)$ spinors, for example, we find
\bea
{\rm tr}_{\rm s}^{\kso(2r+1)}(\mathfrak f_IH^I)^{2m} &=&2^r\times \sum_{m=\sum_i l_i}
\frac{(2m)!}{\prod_i (2l_i)!}\prod_i(\mathfrak f_i/2)^{2l_i}
\eea
The first, $m=2$, reproduces
\bea
{\rm tr}_{\rm s}^{\kso(2r+1)}\cF^{2}&=&2^{r-7}\times\left(16\, {\rm tr}_{\rm def}^{\kso(2r+1)}\cF^{2}\right)\cr\cr
{\rm tr}_{\rm s}^{\kso(2r+1)}\cF^{4}&=& 2^{r-7}\times\left(-8\, {\rm tr}_{\rm def}^{\kso(2r+1)}\cF^{4}+
6\, \left({\rm tr}_{\rm def}^{\kso(2r+1)}\cF^{2}\right)^2\right)
\eea
The same computation works verbatim for the reducible spinor representation
of $\kso(2r)$ with both chiral and anti-chiral spinors combined.

If the individual trace formulae for $\pm$ spinors of $\kso(2r)$ are needed,
we can take the same exercise, as long as $2m<r$. As is evident from the above
Chern classes, the difference between $+$ spinor and $-$ spinor begins to
show starting from $2m=r$. For $2m<r$, the same expansion formula works for
individual $\pm$ spinors, except that one should care to divide the right-hand side by 2. The latter effectively replaces $2^{r-7}$ in front by $2^{r-8}$,
taking into account the half degrees of freedom in the (anti-)chiral  spinor
representations.

\subsection{$\ksp(r=N/2)$}

The case of $\ksp(r)$ is no different. It shares the same $\mathbf q_m(\mathfrak f)$
as the trace of $\cF^{2m}$ in the defining representation, while the adjoint
counterpart is
\bea
{\rm tr}_{\rm adj}^{\ksp(r)}(\mathfrak f_IH^I)^{2m} &=&(4(r-1)+2\times 2^{2m})\times \sum_{i=1}^r (\mathfrak f_i)^{2m} \cr\cr
&&+4\times \sum_{i<j} \sum_{l=1}^{m-1} \frac{(2m)!}{(2l)!(2m-2l)!}
(\mathfrak f_i)^{2l}(\mathfrak f_j)^{2(m-l)}
\eea
where the only difference from $\kso(2r)$ is the additional factor
$2\times 2^{2m}$ from $\pm 2\ke_i$ contributions. Again, starting from the
$\kso(2r)$ results, it is a matter of shifting coefficient of $\mathbf q_m(\mathfrak f)$,
such that $2r$ becomes $2r+2^{2m}$. This results in
\bea
{\rm tr}_{\rm adj}^{\ksp(N/2)}\cF^{2} &=&(N+2) \times {\rm tr}_{\rm def}^{\ksp(N/2)}\cF^{2} \\\cr
{\rm tr}_{\rm adj}^{\ksp(N/2)}\cF^{4} &=&(N+8) \times {\rm tr}_{\rm def}^{\ksp(N/2)}\cF^{4} + 3\times \left( {\rm tr}_{\rm def}^{\ksp(N/2)}\cF^{2}\right)^2
\eea
and so on.

\subsubsection*{Antisymmetric Tensors}

The highest weight of the $n$-th power antisymmetric tensor is $\ke_1+\cdots+\ke_n$.
The weights for $n=2$ are
\bea
\{\pm \ke_i\pm \ke_j\}\cup\{0\}
\eea
which appears identical to the adjoint weights of $\kso(2r=N)$. The hidden
difference is in the degeneracy of the zero weight, which is $r-1$ here
as opposed to $r$ for the $\kso(2r)$ adjoint. However, this difference
does not enter our computation here and the final results for rank 2 antisymmetric tensor
of $\ksp(r)$ are identical to those for the adjoint of $\kso(2r)$,
\bea
{\rm tr}_{\rm \fat[2\fat]}^{\ksp(N/2)}\cF^{2} &=&(N-2) \times {\rm tr}_{\rm def}^{\ksp(N/2)}\cF^{2} \\\cr
{\rm tr}_{\rm \fat[2\fat]}^{\ksp(N/2)}\cF^{4} &=&(N-8) \times {\rm tr}_{\rm def}^{\ksp(N/2)}\cF^{4} + 3\times \left( {\rm tr}_{\rm def}^{\ksp(N/2)}\cF^{2}\right)^2
\eea
where we also used the above fact that the two algebras share the common weights for
the respective defining representations.

The weights for the $n=3$  antisymmetric tensor are the following
\bea
\{\pm \ke_i\pm \ke_j \pm \ke_k\}\cup \{\pm \ke_l\}
\eea
where the latter set of weights are degenerate $r-2$ times.
This leads to
\bea
{\rm tr}_{\rm \fat[3\fat]}^{\ksp(N/2)}\cF^{2} &=&\frac{(N-1)(N-4)}{2} \times {\rm tr}_{\rm def}^{\ksp(N/2)}\cF^{2}\cr\cr
{\rm tr}_{\rm \fat[3\fat]}^{\ksp(N/2)}\cF^{4} &=&\frac{(N-4)(N-13)}{2} \times {\rm tr}_{\rm def}^{\ksp(N/2)}\cF^{4} \cr\cr
 &&+\; 3(N-4)\times \left( {\rm tr}_{\rm def}^{\ksp(N/2)}\cF^{2}\right)^2
 \eea


\providecommand{\href}[2]{#2}\begingroup\raggedright\endgroup

\end{document}